\documentclass[preprint,11pt]{elsarticle}
\usepackage{amsmath}
\usepackage{amssymb}
\usepackage{amsbsy}
\usepackage{amsfonts}
\usepackage{amsopn}
\usepackage{amstext}
\usepackage{graphicx}
\usepackage{amssymb}
\usepackage{amsfonts}
\usepackage{amsmath}
\usepackage{graphicx}
\usepackage[english]{babel}
\usepackage{color}
\usepackage{esint}
\usepackage[dvips]{epsfig}
\usepackage[dvips]{graphicx}
\usepackage{float}
\usepackage{units}
\usepackage{textcomp}
\biboptions{numbers,sort&compress}

\DeclareMathOperator{\e}{e}
\DeclareMathOperator{\x}{x}

\usepackage{hyperref}
\usepackage{slashed}

\journal{Annals of Physics}

\begin{document}

\begin{frontmatter}

\title{Multiresonance modes in sine-Gordon brane models}

\author[IFCE]{W. T. Cruz}
\ead{wilamicruz@gmail.com}

\author[UFC]{R. V. Maluf}
\ead{r.v.maluf@fisica.ufc.br}

\author[UFC]{D. M. Dantas}
\ead{davi@fisica.ufc.br}

\author[UFC]{C. A. S. Almeida}
\ead{carlos@fisica.ufc.br}

\address[UFC]{Universidade Federal do Cear\'a (UFC), Departamento de F\'{i}sica,\\ Campus do Pici, Fortaleza - CE, C.P. 6030, 60455-760 - Brazil}

\address[IFCE]{Instituto Federal de Educa\c{c}\~{a}o, Ci\^{e}ncia e Tecnologia do Cear\'{a} (IFCE),
Campus Juazeiro do Norte, 63040-540 Juazeiro do Norte-Cear\'{a}-Brazil}

\begin{abstract}
In this work, we study the localization of the vector gauge field in two five-dimensional braneworlds generated by scalar fields coupled to gravity. The sine-Gordon like potentials are employed to produce different thick brane setups. A zero mode localized is obtained, and we show the existence of reverberations with the wave solutions indicating a quasi-localized massive mode. More interesting results are achieved when we propose a double sine-Gordon potential to the scalar field. The resulting thick brane shows a more detailed topology with the presence of an internal structure composed by two kinks. The massive spectrum of the gauge field is revalued on this scenario revealing the existence of various resonant modes. Furthermore,  we compute the corrections to Coulomb law coming from these massive KK vector modes in these thick scenarios, where is concluded that the dilaton parameter regulates these corrections.
\end{abstract}

\end{frontmatter}


\section{Introduction}

Over the last years, the physics of extra dimensions aroused great interest with the publication of the seminal works of Lisa Randall and Raman Sundrum about warped geometry \cite{RS}. In such works, our universe is supposed to be represented by a hypersurface (brane) immersed in a larger multidimensional space-time (bulk). The main purpose of the Randall-Sundrum (RS) models is to provide a consistent response to the hierarchy problem and the cosmological constant problem. Despite the initial success, some difficulties involving the field localization and the divergences of geometrical quantities were observed \cite{gremm, de, kehagias}.
In order to remedy the difficulties arising in the RS models it has been proposed the use of  solitons-like topological defects. Such solutions are capable of providing an internal structure to the brane and thereby improve the field location mechanism and also to prevent singularities in the scalar curvature, in the Schr\"{o}dinger quantum analogue potential and in physical quantities of some other models \citep{defects1, defects2, aplications,fase}. One of the advantages of this proposal is to provide a method for dynamically generating the brane through scalar field coupled to gravity in the presence of the $\phi^4$ potential. The solutions obtained in these scenarios give rise to different kinds of topological defects and enable the development of new and interesting phenomena. For example, the presence of resonant states resulting from the interaction of the fields with the potential responsible for generating the brane.

Several studies of field localization were carried out in thick braneworlds with an extra dimension. Such studies were initially aimed to explain how the internal structure of the defects involving kink solutions obtained from the potential $\phi^4$ can interfere with localization of different fields on the brane \cite{nosso1, nosso2, nosso3, nosso4}. As an example, we can cite the Bloch Brane models generated by one or more scalar fields \cite{nosso6, nosso7, nosso8}, in which the localization of tensorial and spinorial fields and the search for resonant states were the subject of several works \citep{nosso7, carlos, chineses6, chineses7}.

Another way to get solitonic solutions is to consider potential of sine-Gordon (SG) type to construct the brane model, as studied in Refs. \cite{sg1,sg2,sg3,sg4}. More recently, we consider the sine-Gordon and the double sine-Gordon (DSG) potentials in order to study the location of gravity.

New solutions for thick branes taking into account SG potential with broken symmetry $Z_2$ was carried out in \cite{sg5} where the authors verify the stability and confinement effects for these braneworld models. However, the important issue about the field trapping and the existence of quasi-localized massive modes for the other fields present in the Standard Model have not been properly addressed in the context approached by us in Ref. \cite{annals}. Thus, in this paper we propose to analyse the localization of the vector field on the brane model generated by the potentials SG and DSG. The localization mechanism adopted by us and widely used in the literature consists of coupling directly into the five-dimensional Lagrangian density the dilaton field with the usual Maxwell kinetic term. In order to analyse the influence of massive vector  modes in the Coulomb law, we also perform the fermion field localization in the SG and the DSG branes. This method of massless spin $1/2$ confinement is very often in 5D models \cite{nosso3, carlos, chineses6, chineses7, sg1}.

This work is outlined as follows: In section \ref{rev_sg_sce} we introduce a braneworld generated by the SG potential and describe the localization of the vector field zero mode and massive spectrum. Next, in section \ref{dil_bb}, we include the dilaton field on the SG brane setup and reappraise the gauge field on this new scenario. In section \ref{mass_mode_dsg} we introduce the DSG potential to the scalar field and obtain new results over the gauge field massive modes. Further, in the section \ref{sec-coulmb}, we compute the slight deviation in the Coulomb law due these massive vector modes and due the massless fermion left mode. Finally, in section \ref{conclu}, we present comments about our results.

\section{The Sine-Gordon Brane \label{rev_sg_sce}}
We consider a five-dimensional model described by one scalar field coupled to gravity, to generate a thick brane. In this scenario, we assume that the scalar field depends only on the extra dimension $y$, and that the spacetime is asymptotic $AdS_{5}$ described by the metric $ds^{2}=e^{2A(y)}\eta_{\mu\nu}dx^{\mu}dx^{\nu}+dy^{2}$. Also, we used throughout this work the signature $(-,+,+,+)$ for the flat metric $\eta_{\mu\nu}$ with $\mu,\nu=0,1..3$. The function $A(y)$ characterizes the warp factor, and it will be fixed with the particular choice of potential.

With these preliminaries at hand, we can write the action of the form\begin{equation}\label{eq:action}
S=\int d^{5}x\sqrt{-G}\Big[\frac{1}{4}R-\frac{1}{2}\partial_{\mu}\phi\partial^{\mu }\phi-V(\phi)\Bigr],
\end{equation} in which $R$ is the scalar curvature. The equations of motion read off from this action are \begin{equation}
\phi'^{2}-2V(\phi)  =  6A'^{2}\label{eq:eqmov1}
\end{equation}
\begin{equation}
\phi'^{2}+2V(\phi)  =  -6A'^{2}-3A''\label{eq:eqmov2}
\end{equation}
\begin{equation}
\phi''+4A^{\prime}\phi'  =  \frac{\partial V}{\partial \phi},
\label{eq:eqmov3}
\end{equation}
where prime stands for derivative with respect to $y$.

Now, we take the potential $V$ on the form
\begin{equation}\label{eq:vsup} V(\phi)=\frac{1}{8}\left(\frac{\partial W}{\partial\phi}\right)^{2}-\frac{1}{3}W^{2},
\end{equation}such that the Eqs. \eqref{eq:eqmov1} - \eqref{eq:eqmov3} result in the following first order equations:\begin{equation}\label{eq:firstorder}\phi^{\prime}=\frac{1}{2}\frac{\partial W}{\partial\phi},
\end{equation}
\begin{equation} A^{\prime}=-\frac{1}{3}W.
\label{eq:A}
\end{equation}

It should be mentioned that the method used to linearize the above equations of motion is well known in the literature, and it consists of writing the potential $V(\phi)$ as a function of a superpotential $W(\phi)$. For further details, see
reviews in the Refs. \cite{bazeia1,bazeia2,bazeia3,bazeia4,shif,alonso,de, skenderis, cvetic} and references therein. 

Now, we make the connection to the sine-Gordon (SG) model taking the superpotential in the form
\begin{equation}\label{eq:supot}
W(\phi)=3 b c \sin\left(\sqrt{\frac{2}{3 b}\phi}\right),
\end{equation}which implies the potential
\begin{equation}\label{eq:sgpot}
V(\phi)= \frac{3}{8}b c^2 \left[1- 4 b+ \left(1+ 4b\right) \cos\left(\sqrt{\frac{8}{3b}}\phi\right)\right].
\end{equation}

The solution to the first order equation \eqref{eq:firstorder} is obtained immediately by taking the superpotential defined in \eqref{eq:supot}, such that
\begin{equation}
\phi(y)=\sqrt{6b}\arctan\lbrace\tanh\left[\frac{1}{6}\left(3 c y+ \sqrt{\frac{6}{b}}C_1\right)\right]\rbrace \label{eq:phi}.
\end{equation}
Now, using the equations \eqref{eq:eqmov1}-\eqref{eq:eqmov2}, we can obtain the following warp factor
\begin{equation}\label{eq:warpfactor}
e^{2A(y)}= {\cosh\left[c y + \frac{\sqrt{\frac{2}{3}} C_
1}{\sqrt{b}}\right]}^{-2 b}.
\end{equation}

Hereinafter, the constant $C_1$ is assumed to be zero to make the defect centred at $y = 0$. The remaining constants, $b$ and $c$, adjust the $\phi$ value when $y \rightarrow\pm\infty$ and control the thickness of the defect, respectively.

\subsection{Gauge field zero mode \label{gau_f_ze_mode}}

Now, we will study the zero mode of the gauge field on the thick brane set up by the SG potential obtained in the previous section. Consider the action that describes a vector field coupled to gravity in five dimensions:
\begin{equation}\label{action1}
S\sim\int d^{5}x\sqrt{-G}F_{MN}F^{MN},
\end{equation}
where $F_{MN}=\partial_M A_N -\partial_N A_M$ with  $M,N=1,2,...,5$.

The equation of motion for $A_{\mu}$ can be brought in the form
\begin{equation}\label{eq:zero_gauge}
-\frac{d^{2}U(y)}{dy^{2}}-2 A'(y)\frac{dU(y)}{dy}=m^{2}e^{-2A(y)}U(y),
\end{equation}with the convenient gauge choice $\partial_\mu A^\mu=A_y=0$,  along with the ansatz
\begin{equation}\label{sep}
A_\mu(x,y)=a_\mu (0) e^{i p\centerdot x}U(y), \,\,\,\,\,  p^2=-m^2.
\end{equation}
We note that the space-time coordinates of the observable universe are defined by $x$, while the function $U(y)$ condenses all the information of the extra dimension on the behaviour of the vector field. Given the Eq. \eqref{sep} we can rewrite the action \eqref{action1} in the form
\begin{equation}\label{acao}
S\thicksim\int d^{5}x \sqrt{-G}~ F_{MN}F^{MN}= \int dy U(y)^2 \int d^4 xf_{\mu\nu}(x)f^{\mu\nu}(x).
\end{equation}Thus, we need to investigate the effects of solution $U(y)$ due to \eqref{eq:zero_gauge} to evaluate the existence of vector modes in the brane.

The effective action \eqref{acao} immediately tells us that the warp factor is absent and its exponential suppression effect no longer applies.  The existence of a normalized solution that makes finite the effective action is ensured solely by the solutions of the equation of motion for the vector field in the extra dimension. For the zero mode with $m=0$, a possible solution is $U(y)=constant$, which implies a divergent effective action. Thus, this particular solution does not allow the existence of a gauge field zero mode on this background.

\begin{figure}
\centering
\includegraphics[width=0.5\textwidth]{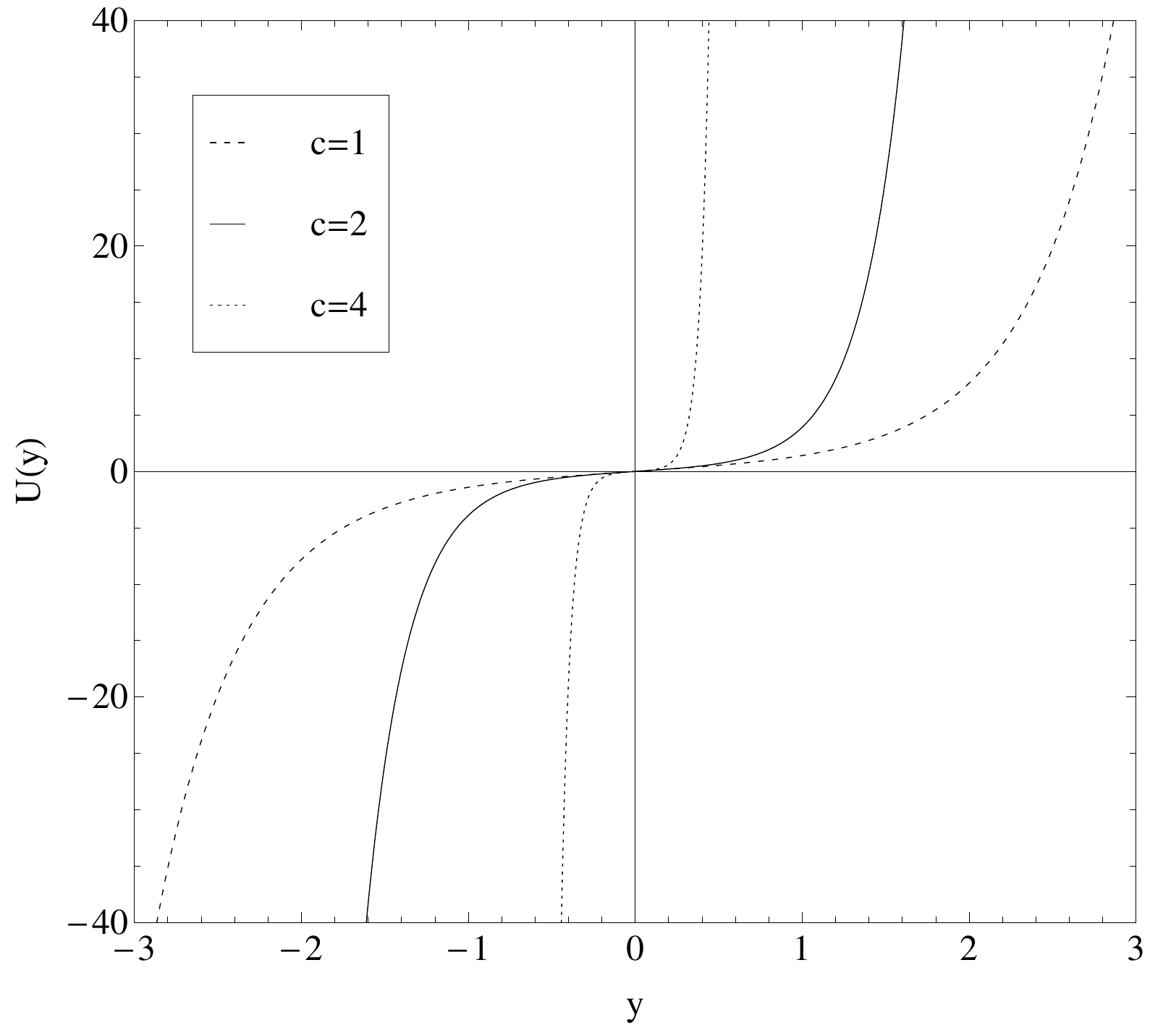}
\caption{{ Plots of the solution $U(y)$ in equation (\ref{U0_2}).}}\label{sgg1}
\end{figure}

A general solution to equation \eqref{eq:zero_gauge} with $m=0$ has the form
\begin{equation}\label{U0}
U(y)=\int_{y_0}^y e^{-2A}dy'.
\end{equation}
Taking $A(y)$ from (\ref{eq:warpfactor}), setting $y_0=0$ and $b=1$,  the solution above assumes the form 
\begin{equation}\label{U0_2}
U(y)=\frac{2cy+\sinh(2cy)}{4c}
\end{equation}
However, due to the shape of the function $A(y)$, the solution $U(y)$ diverges for any value of $c$. In Fig. \ref{sgg1} we plot $U(y)$ as a function of $y$ for different values of $c$. We note that when we increase the thickness of the kink, the zero mode solution diverges more rapidly, which renders the effective action infinite. Therefore, the existence of a localized zero mode with this solution is not guaranteed.

\subsection{Massive modes  \label{mass_mode_res}}

Let us now focus our attention on the Kaluza-Klein massive modes. Considering the case $m\neq0$ in equation \eqref{eq:zero_gauge}, it is convenient to turn it into a Schr\"{o}dinger-like equation  through the prescription
\begin{equation}\label{t}
y\rightarrow z=f(y),\,\,\,\,\,\,f'(y)=e^{-A}\,\,\,\,\,\, U(y)=e^{-\frac{1}{2} A}\overline{U}(z).
\end{equation}
Thus, we get an equation in the form
\begin{equation}\label{eq:schgauge}
\left\{-\frac{d^2}{dz^2}+{V}(z)\right\}\overline{U}(z)=m^2\overline{U}(z),
\end{equation}where $V(z)= -\frac{7}{4}\dot{A}^2+\frac{1}{2}\ddot{A}$ is the one-particle potential, and the dots represent derivatives with respect to new coordinate $z$.
Applying the transformations defined above, we can write the warp factor (\ref{eq:warpfactor}) in the form
\begin{equation}
A(z)=-\ln\left[\sqrt{(z^2 c^2 + 1)}\right],
\end{equation}where we set $b=1$ for simplicity, such that the potential becomes expressed as
\begin{equation}\label{eq:potz}
V(z)=-\frac{c^2 (2 + 5 c^2 z^2)}{4(1 + c^2 z^2)^2}.
\end{equation} 

This equation cannot be written in the form corresponding to supersymmetric quantum mechanics $Q^{\dag}\,Q\,\overline{U}(z)=-m^2\overline{U}(z)$. Moreover, the zero mode of the equation  (\ref{eq:schgauge}) is not bound on the brane, as showed in figure (\ref{sgg4_2}). Thus, we cannot exclude the possibility of tachyonic states \cite{dionisio_tachyons, chineses4b}.

\begin{figure}
\centering
\includegraphics[width=0.99\textwidth]{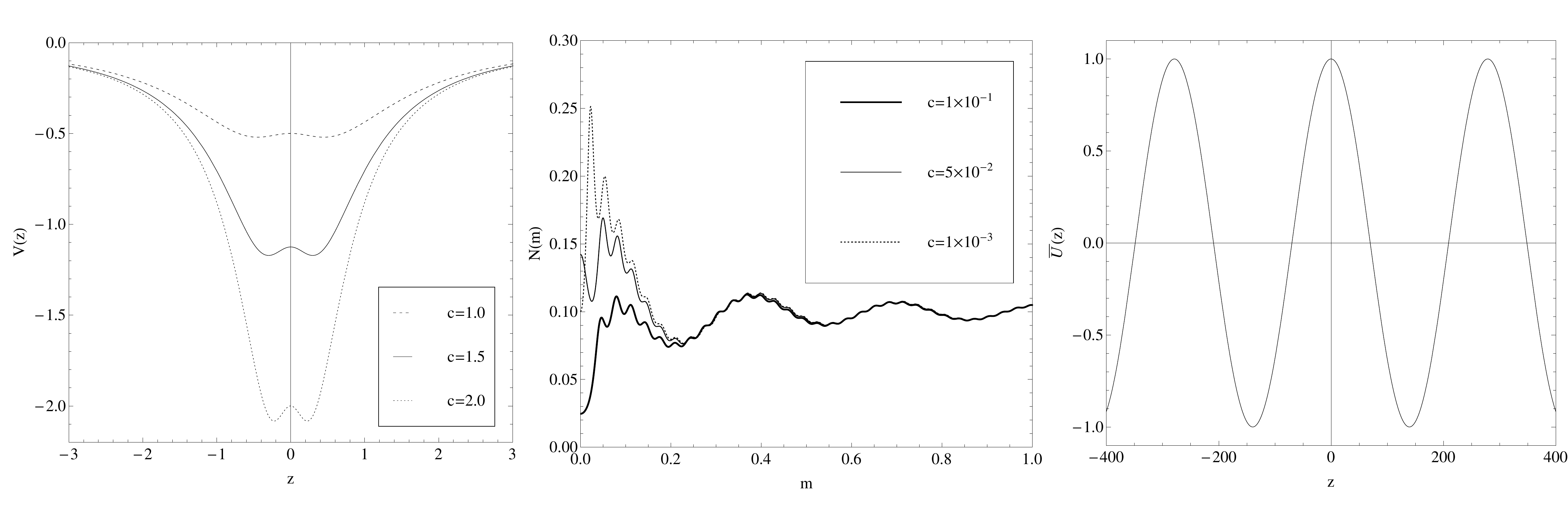}
\caption{ Plots of the potential (\ref{eq:potz}) (left), $N(m)$ (center) (\ref{eq:rel_prob}) and the solution $\overline{U}(z)$ for $m=0.02252$.}\label{sgg3}
\end{figure}

The coupling of the massive modes with the matter on the brane is established in terms of the solution to the Schr\"{o}dinger-like equation at $z=0$. The solutions to $U(z)$ acquire plane wave structure when $m^2\gg V_{max}$ because the potential  represents only a small perturbation. Thus, if there are resonant modes we expect that they must emerge with $m^2\leq V_{max}$. The structure of the resulting potential (\ref{eq:potz}), which we show in Fig. \ref{sgg3} suggests that there are no reverberations with the wave solutions. As we have noted, the potential tends to infinity with negative values, and the maximum values are also negative. In addition, we note that the potential does not have the usual volcano-like structure which has been obtained in thick brane models with localized zero modes as well as gauge field resonant modes.

In general, resonant mode are reverberations of the wave solutions near the bounce $z=0$. These echoes are characterized by solutions to $U(z)$ around $z=0$ with large amplitudes in comparison with its values far from the defect. To seek for such structures we adopt a largely used method based on the relative probability $N(m)$ \cite{chineses1,chineses2,chineses3, chineses4, chineses5} given by
\begin{equation}\label{eq:rel_prob}
N(m)=\frac{\int_{-z_{b}}^{+z_{b}}|\overline{U}_{m}(z)|^2 dz}{\int_{-z_{max}}^{+z_{max}}|\overline{U}_{m}(z)|^2dz}.
\end{equation}
From the Eq. (\ref{eq:schgauge}) we can consider $\zeta|\overline{U}_{m}(z)|^2$  as the probability for finding the mode at the position $z$, where $\zeta$ is a normalization constant. Thus, resonant modes will be identified by peaks in $N(m)$.

To evaluate the relative probability, we take a narrow integration range around the brane $-z_b<z<z_b$ with $z_b=0.1z_{max}$ \cite{chineses3}, and the massive modes considered inside a box with borders $|z|=z_{max}$ far from the turning points of the potential \cite{chineses1, chineses4}. The choice of the integration interval $z_b=0.1z_{max}$ does not interfere with the values of the masses to the possible resonance modes \cite{nosso8}. A symmetric brane should result in even and odd solutions of $\bar{U}(z)$ \cite{chineses4b}. Hence we use the initial conditions $\overline{U}(0)=1$ and $\overline{U}'(0)=0$ for the even solutions and the conditions $\overline{U}(0)=0$ and $\overline{U}'(0)=1$ for the odd solutions \cite{chineses4b}.


In fact, there are some peaks in $N(m)$ as showed in the Fig.  \ref{sgg3}. However, when we take the mass corresponding to the higher value of $N(m)$ and evaluate the corresponding $\overline{U}(z)$ from (\ref{eq:schgauge}), a plane wave solution is obtained. This behaviour confirms that, besides the fact the zero mode is not localized, the potential (\ref{eq:potz}) supports no reverberations and there are no gauge field massive modes highly coupled to the brane in the present setup. Finally, examining the result of Eq. (\ref{eq:rel_prob}) at $m=0$, no disproportional value was found, confirming that there are no evidences of bound state localization.

\section{Sine-Gordon brane with the dilaton field \label{dil_bb}}

In this section, we will introduce a new scalar field in our model, the dilaton field, which will play a key role in solving the localization problem. Besides, there are very recent phenomenological perspectives showing promising results for the dilaton field in the LHC \cite{dilatonlhc1, dilatonlhc2, dilatonlhc3}. The new thick brane setup is described by the action \cite{kehagias, mk1, nosso1,mk2, nosso2}
\begin{equation}\label{acdil}
\mathcal{S}=\int d^5x\sqrt{-G}\left(\frac{1}{4}R-\frac{1}{2}(\partial\phi)^2-\frac{1}{2}(\partial\pi)^2-V(\phi,\pi)\right),
\end{equation} and our ansatz for the metric becomes \cite{kehagias}
\begin{equation}
ds^2=e^{2A(y)}\eta_{\mu\nu}dx^{\mu}dx^{\nu}+e^{2B(y)}dy^2.
\end{equation}
Thus, the equations of motion obtained from this modified action can be written as
\begin{eqnarray}\label{eq-of-motion2}
\phi'^{2}+\pi'^{2}-2e^{2B}V&=&6A'^{2}\\\nonumber
\phi'^{2}+\pi'^{2}+2e^{2B}V&=&-6A'^{2}+3A'B'-3A''\\\nonumber
\gamma''+(4A'-B')\gamma'&=& e^{2B}\partial _{\gamma}V, \,\,\,\,\, \gamma=\phi,\pi.
\end{eqnarray}
To achieve first-order differential equations, we employ the same superpotential method implemented by  equation  \eqref{eq:vsup}. In order to preserve the structure of the SG brane, we consider the following potential function \cite{kehagias}
\begin{equation}\label{eq:pot_dil}
V(\phi,\pi)=e^{\pi \sqrt{\frac{2}{3}}}\left[\frac{1}{8}\left(\frac{\partial W}{\partial \phi}\right)^2 -\frac{5}{16}W^2\right].
\end{equation}

\begin{figure}
\centering
\includegraphics[width=1.00\textwidth]{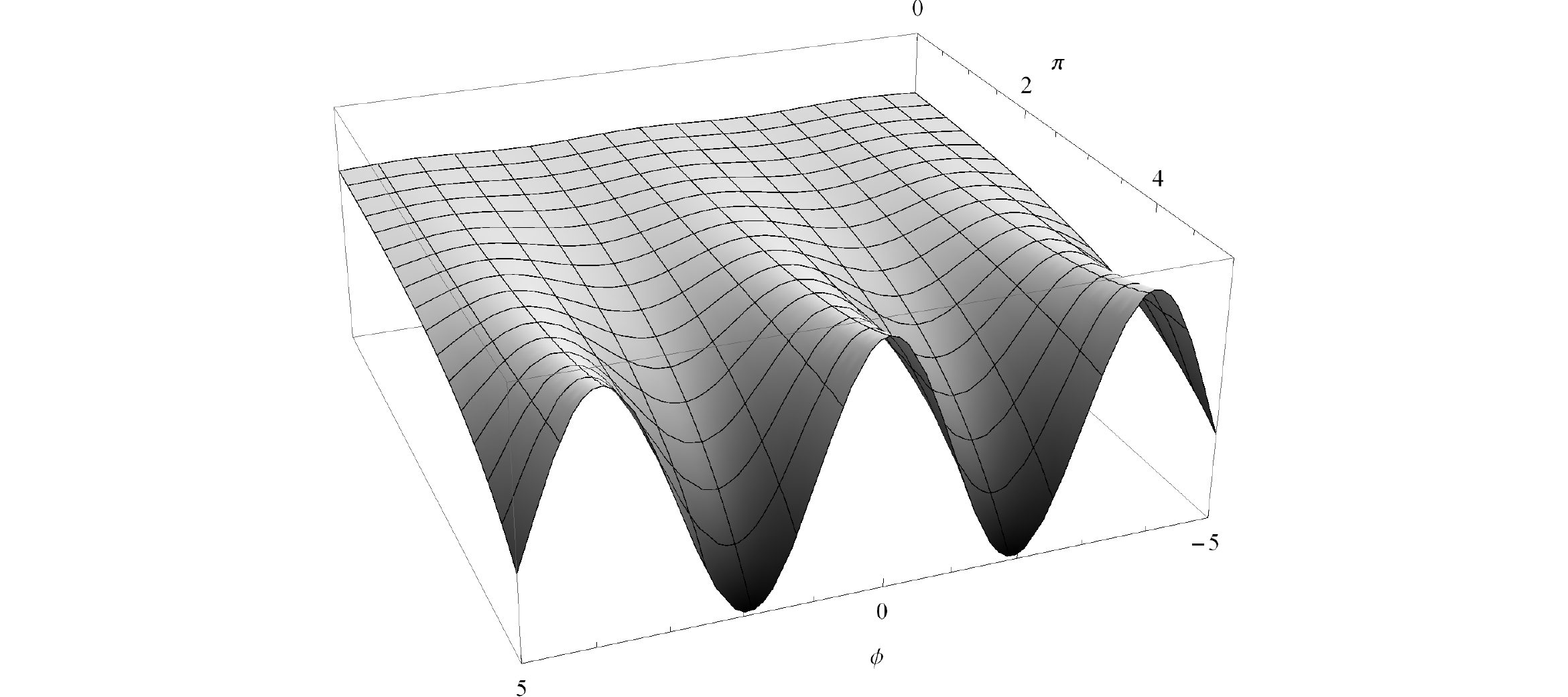}
\caption{{ Plots of the potential $V(\phi,\pi)$ of equation (\ref{eq:pot_dil2}).}}\label{sgg2}
\end{figure}

It is immediately verified that the equations of motion \eqref{eq-of-motion2} are solved when we consider the following set of first-order differential equations:
\begin{eqnarray}\label{eq:singularidade1}
\phi^{\prime}=\frac12\,\frac{\partial W}{\partial\phi},\\\nonumber
\pi=-\sqrt{\frac{3}{8}}A,\\\nonumber
B=-\frac{\pi}{2}\sqrt{\frac{2}{3}}=\frac{A}{4},\\\nonumber
A'=-\frac{W}{3}.
\end{eqnarray} The solutions for $\phi$ and $A(y)$ shown in equations \eqref{eq:phi} and \eqref{eq:warpfactor} remain valid for the choice of superpotential as in \eqref{eq:supot}. However, the potential is modified according to the equation \eqref{eq:pot_dil}, which leads to the result
\begin{equation}\label{eq:pot_dil2}
V(\phi,\pi)=e^{\pi \sqrt{\frac{2}{3}}}\left[\frac{3}{32}b c^2 \left(4- 15 b+ \left(4+ 15b\right) \cos\left(\sqrt{\frac{8}{3b}}\phi\right)\right)\right].
\end{equation}This is plotted in Fig. \ref{sgg2}.

Taking into account this new background, we define the kinetic term for the gauge field coupled to the dilaton by the effective action \cite{kehagias, mk1, nosso1,mk2, nosso2} 
\begin{equation}
S\sim\int d^{5}x\sqrt{-G}~e^{- 2\lambda \pi\sqrt{\frac{2}{3}}}F_{MN}F^{MN},
\end{equation}
where $\lambda$ represents a parameter that regulates the dilaton coupling.

As before, we assume $\partial_\mu A^\mu=A_y=0$ and the ansatz $A_\mu(x,y)=a^\mu (0) e^{i p\centerdot x}U(y)$ with $p^2=-m^2$ such that the equation for gauge field component that is a function only of the extra dimension is given by
\begin{eqnarray}\label{zero2dil}
-\frac{d^{2}U(y)}{dy^{2}}-\left[2 A'-B'-2\lambda \pi'\sqrt{\frac{2}{3}}\right]\frac{dU(y)}{dy}=m^{2}e^{2(B-A)}U(y).
\end{eqnarray}
One sees from the above equation that $U(y)=constant$ is a solution for the case $m=0$. This result makes the gauge field zero modes localized on the brane. Indeed, we take the effective action and decompose it in the following way
\begin{equation}
\int d^{5}x \sqrt{-G}~e^{- \lambda \pi\sqrt{\frac{8}{3}}}F_{MN}F^{MN}= \int dy U(y)^2 e^{A(y)(\frac{1}{4}+\lambda)} \int d^4 xf_{\mu\nu}(x)f^{\mu\nu}(x),
\end{equation}
where the proportionality relation between the functions $\pi(y)$ and $A(y)$ was used.

Note that for  $\lambda\geqslant-1/4$ the effective action is finite and the solution $U(y)$ is normalized. The effect of dilaton coupling is to restrict  the range of the gauge field in the extra dimension, a key feature when we look for localized states. Hence, we conclude that the dilatonic sine-Gordon brane does support the existence of gauge field zero modes.

\subsection{Massive modes and resonances \label{mass_mode_res}}

\begin{figure}
\centering
\includegraphics[width=0.95\textwidth]{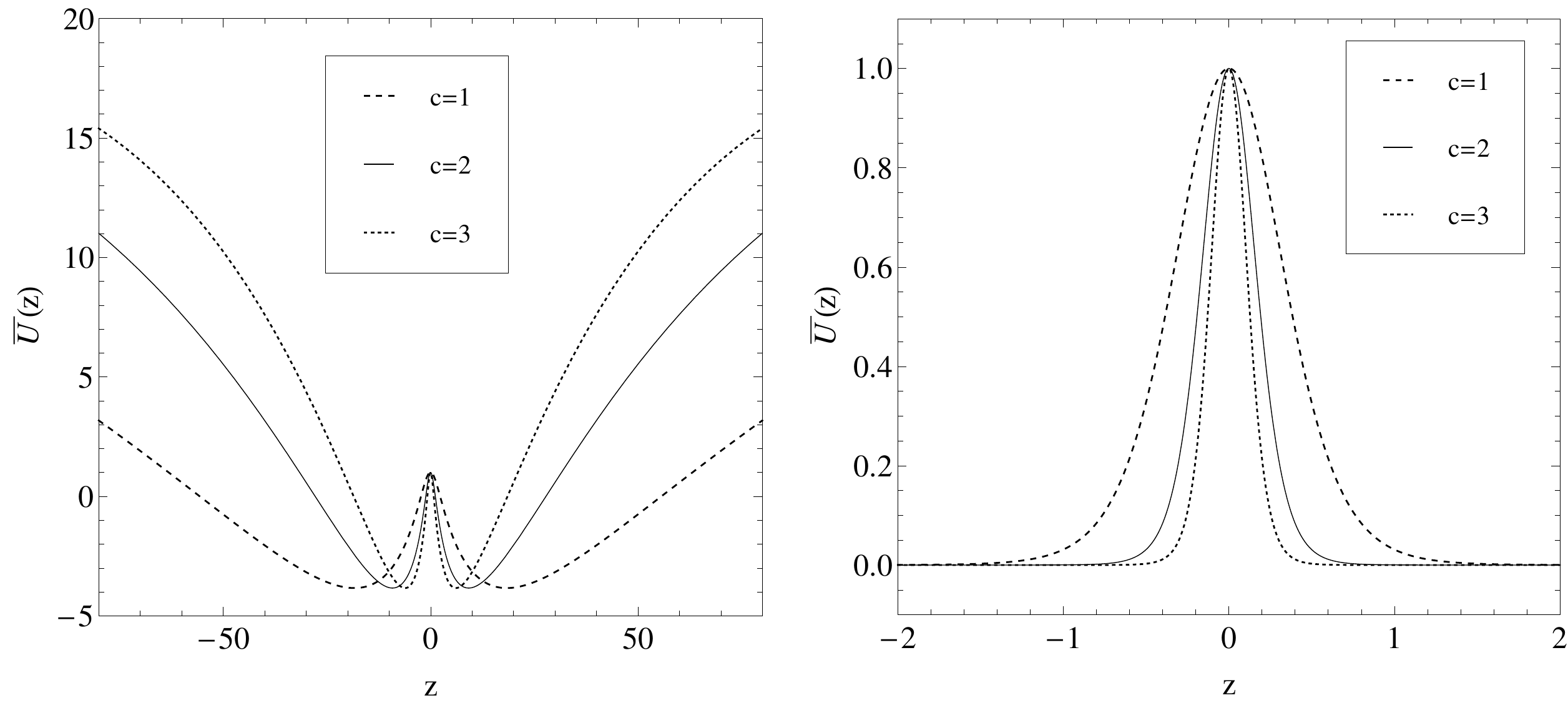}
\caption{Zero mode solutions of the equations (\ref{eq:schgauge}) (left) and (\ref{schgauge}) (right). The figure on the right side correspond to the gauge field with the dilaton coupling.}\label{sgg4_2}
\end{figure}

Now, we will focus our attention on the issue about the existence of the massive modes quasi-localized. To analyse the Kaluza-Klein massive spectrum, we take the equation \eqref{zero2dil} to $m\neq0$ and leaves it into a Schr\"{o}dinger-like equation. This is accomplished by means of the prescription
\begin{equation}\label{t}
y\rightarrow z=f(y),\,\,\,\,\,\,f'(y)=e^{-\frac{3}{4}A}\,\,\,\,\,\, U(y)=e^{-\gamma A}\overline{U}(z),
\end{equation}
where $\gamma=(\lambda+1)/2$. Thus, we find the following
Schr\"{o}dinger-like equation
\begin{equation}\label{schgauge}
\left\{-\frac{d^2}{dz^2}+{V}(z)\right\}\overline{U}(z)=m^2\overline{U}(z),
\end{equation}
with the one-particle potential $V(z)=\gamma (\gamma\dot{A}^2+\ddot{A})$. Here dot means derivative with respect to $z$.

After the transformations above, setting $b=\frac{4}{3}$ to simplify our results, the $A(z)$ function assumes the following structure
\begin{equation}
A(z)=-\ln\left[\sqrt{(z^2 c^2 + 1)}\right],
\end{equation}
with the potential given by
\begin{equation}\label{eq:potsch}
V(z)=\frac{c^2 \gamma (-1 + c^2 z^2 (1 + \gamma))}{(1 + c^2 z^2)^2}.
\end{equation}

\begin{figure}
\centering
\includegraphics[width=0.99\textwidth]{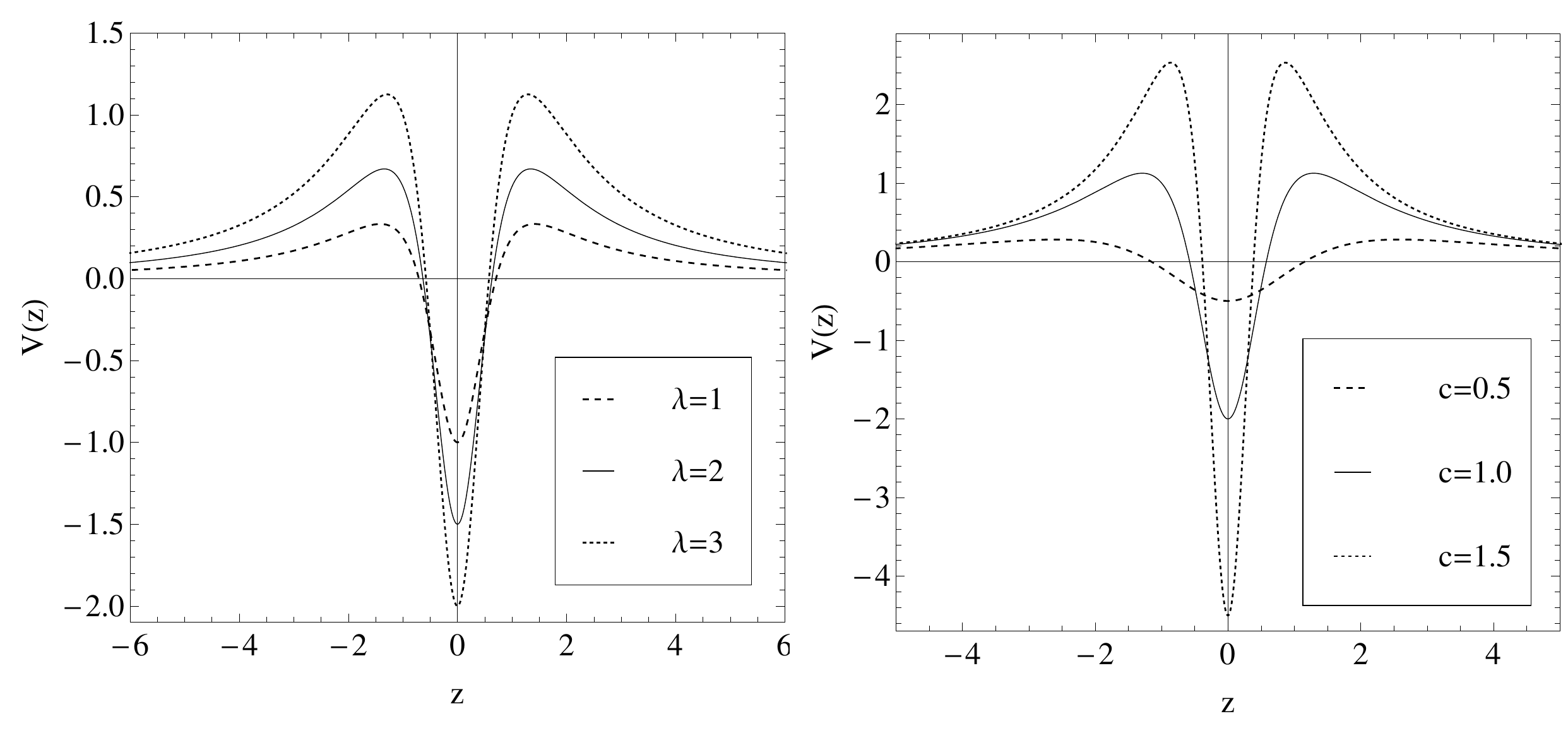}
\caption{{ Plots of the potential $V(\phi,\pi)$ in equation (\ref{eq:pot_dil2}).}}\label{sgg4}
\end{figure}

Equation (\ref{schgauge}) can also be written as the form corresponding to supersymmetric quantum mechanics
\begin{equation}\label{eq:susy_qm_gauge}
Q^{\dag} \, Q \,
\overline{U}(z)=\left\{\frac{d}{dz}+\gamma\dot{A}\right\}\left\{\frac{d}{dz}-\gamma\dot{A}\right\}\overline{U}(z)=-m^2\overline{U}(z).
\end{equation}
The zero mode of the equation (\ref{schgauge}) is bound on the brane as plotted on the Fig. (\ref{sgg4_2}). The ground state has the lowest mass square $m^2_0 = 0$ and there is no tachyonic gauge field mode.

For the purpose of knowing the features of the wave solutions to the gauge field, we must understand the behaviour of the potential (\ref{eq:potsch}). In fact, the existence of reverberations in the massive spectrum is connected with the structure of the volcano-like potential. The maxima of the potential must be separated enough to allow the plane wave to resonate inside the well. The potential is plotted in Fig. \ref{sgg4} varying the thickness of the bounce and the dilaton coupling. Increasing the value of the constant $c$ makes deeper and thinner the potential well. However, the variation of $\lambda$ do not change significantly the thickness of the well but can raise the maxima of  $V(z)$. Reverberations may occur for $m^2\leq V_{max}$, so,  increasing the dilaton coupling, we can expand the mass spectra where they can be found.

\begin{figure}
\centering
\includegraphics[width=1\textwidth]{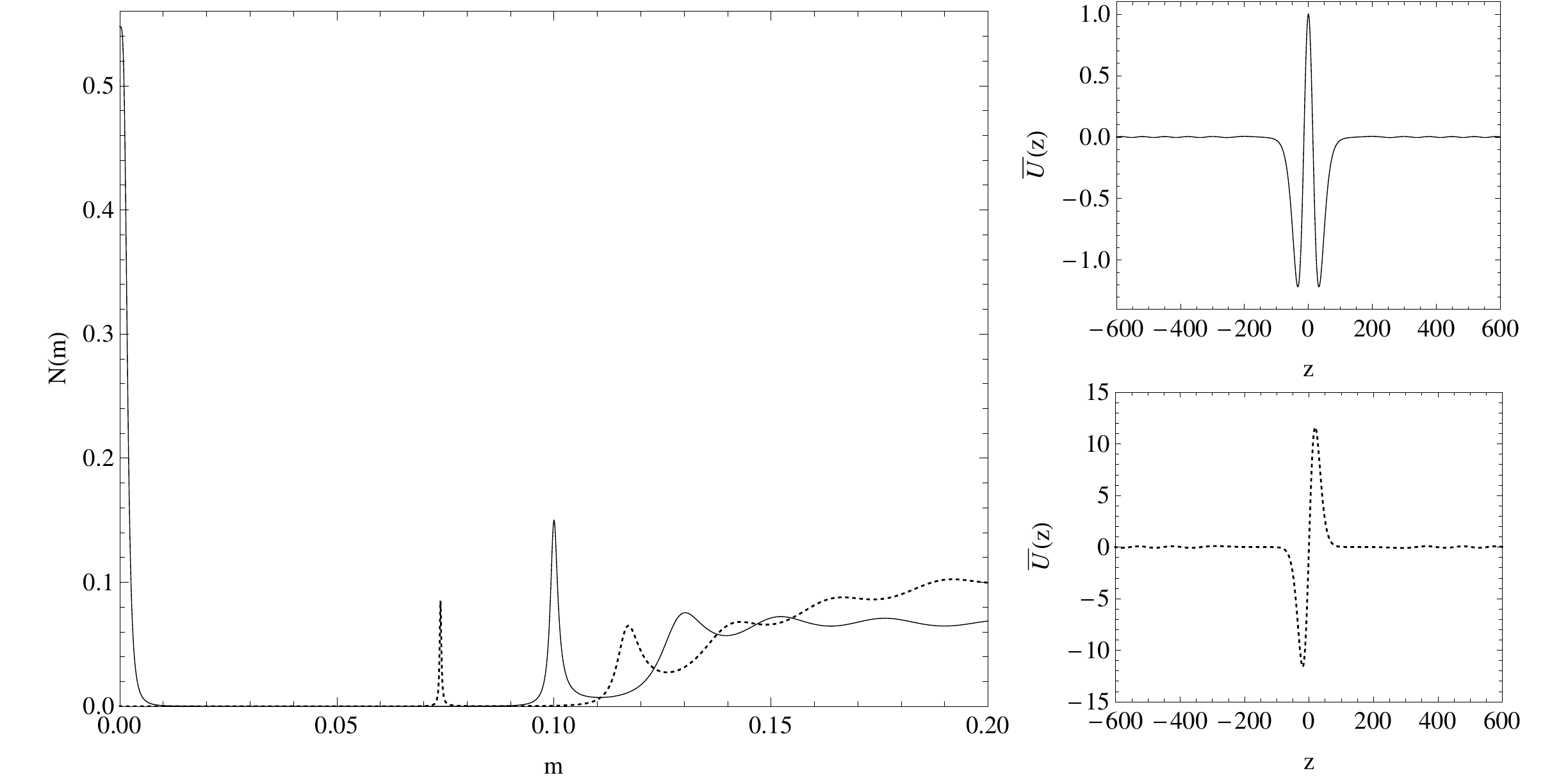}
\caption{
N(m) function (\ref{eq:rel_prob}) and the corresponding solutions $\overline{U}(z)$ of the two resonances found for even (solid line) and odd (points) solutions. Results obtained from the equation (\ref{eq:susy_qm_gauge}) with $c = 0.012$ and $\lambda=40$ .}\label{sgg5}
\end{figure}

Combining  the features above, we can conclude that the best scenario to achieve resonant states is obtained making the defect more thicker (reducing $c$) and increasing the dilaton influence. To verify this thesis, we reset the relative probability function (\ref{eq:rel_prob}) with the solutions of (\ref{eq:susy_qm_gauge}). The result is showed in Fig. \ref{sgg5}, where we note drastic changes in the results in comparison with the Fig. \ref{sgg2}.  Now, we have three peaks on the relative probability function. The peak at $m=0$ indicates the localization of the massless mode. The following peaks, at $m=0.0738861$ and $m=0.0999432$, are odd and even resonances. This can be confirmed by the solution to $\overline{U}(z)$ with the mass value, that is also plotted in Fig. \ref{sgg5}.

\section{Double sine-Gordon Brane \label{mass_mode_dsg}}

Now, we can construct a new thick brane scenario from an extension of the basic SG model. The double sine-Gordon potential (DSG) provides two-kink solutions that can be used to model thick branes \citep{dsg3,dsg4,dsg5,aplications,dsg_ap2,dsg_ap3}. Previous works have showed that splitting branes with internal structure, generated by two-kink defects, can foment the presence of resonant states \cite{nosso6, nosso7, nosso8}. Based on these results we now propose to construct a brane model generated by a DSG potential with the presence of the dilaton field. The DSG model provides a new behaviour about the nucleus of the brane. There is a flat region at the brane location and a splitting on the matter energy density, as described in the work \cite{annals}. 

Starting from the action (\ref{acdil}), we propose the new superpotential
\begin{eqnarray}\label{eq:wdsg2}
\mathcal{W}(\Phi)= 4 \cos(\Phi/2) \sqrt{a + 2\big(1+ \cos(\Phi)\big)}\nonumber \\ + 2 a \ln\left[2 \cos(\Phi/2) + \sqrt{a + 2\big(1+ \cos(\Phi)\big)}\right].
\end{eqnarray}

We adopt the function above because it provides, in flat space-time, a DSG potential of $\Phi$ with the potential defined as $\mathcal{V}(\Phi)=1/8(d \mathcal{W}/d \mathcal{\phi})$. Now we need to consider a curved space-time and the dilaton coupling. 
Therefore, to maintain the first order equations (\ref{eq:singularidade1}) we write the potential as (\ref{eq:pot_dil}) with the new superpotential (\ref{eq:wdsg2}), that results in
\begin{eqnarray}\label{eq:v_dsg}
\mathcal{V}(\Phi,\pi)=e^{\pi \sqrt{\frac{2}{3}}}\Bigg[ 1 -\cos(2\Phi) + a\big(1 - \cos(\Phi)\big) -\nonumber \\ \frac{5}{4}\Bigg[2 \cos(\Phi/2) \sqrt{a + 2\big(1+ \cos(\Phi)\big)} +\nonumber \\
a \ln\left(2 \cos(\Phi/2) + \sqrt{a + 2\big(1+ \cos(\Phi)\big)}\right)\Bigg]^2\Bigg].
\end{eqnarray}
\begin{figure}
\centering
\includegraphics[width=0.99\textwidth]{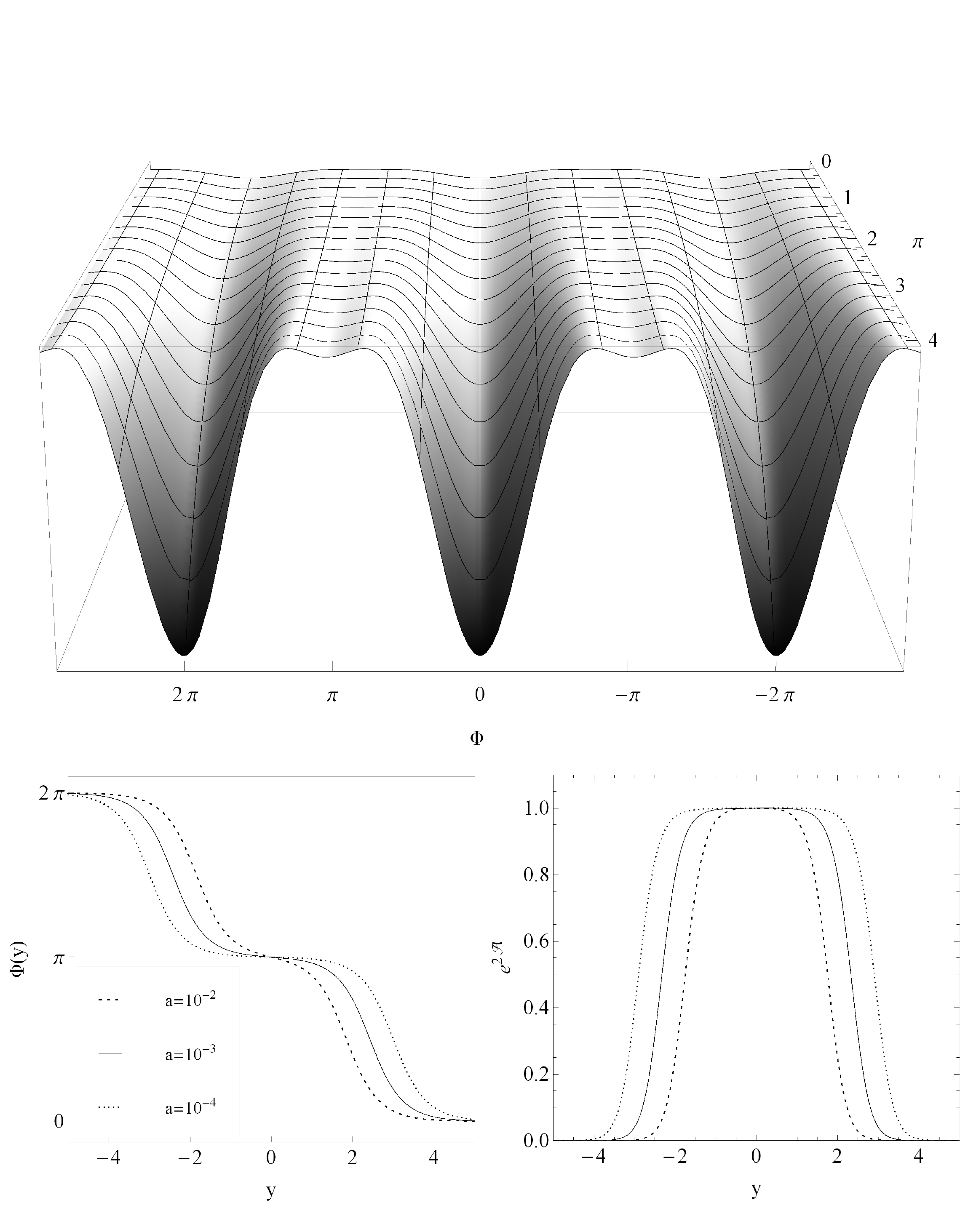}
\caption{{ Plots of the potential $\mathcal{V}(\Phi,\pi)$ (\ref{eq:v_dsg}), the corresponding $\Phi(y)$ and the warp factor.}}\label{sgg6}
\end{figure}

The structure of  $\mathcal{V}(\Phi,\pi)$ is plotted in the Fig. \ref{sgg6}. In terms of $\Phi$ the DSG potential has a absolute minimum at $\Phi=0$ and we have the raising of a relative minima at $\Phi=\pm\pi$. As showed in previous works \cite{dsg1,dsg2,dsg3,dsg4}, the changing in the original vacua of the SG model modifies the scalar field solution. The new bounce solution interpolates between the two vacua ($ 0$ and $2\pi$) with a transient state at $\pi$ \cite{dsg1,dsg2,dsg3,dsg4}.

The new solution  to the scalar field on the DSG scenario must be obtained from the first-order equation $\Phi^{\prime}=\frac{1}{2}\frac{\partial \mathcal{W}}{\partial\Phi}$. Due to the more complex structure of $\mathcal{W}(\Phi)$ than in the SG model we are unable to evaluate $\Phi$ analytically. Thus, we plot the numerical solution for $\Phi$ in Fig. \ref{sgg6}, where we note that the constant $a$ in $\mathcal{W}(\Phi)$ controls the thickness of the brane. This new bounce solution interpolates between the vacua at $0$ and $2\pi$ with a intermediate sector related to the intermediate minimum of the potential at $\pi$. The same behaviour was found in structures, called two-kink solutions, from a deformation of $\Phi^4$ potentials \cite{aplications,dionisio_tachyons,nosso2}. 

The solution for the warp factor is obtained from equation $\mathcal{A}'(y)=-1/3\mathcal{W}(\Phi)$. Based on numerical data for $\Phi$, we construct the warp factor, that is also plotted in Fig. \ref{sgg6}.

The gauge field zero mode localization is guaranteed by the dilaton coupling, however the massive spectra must be changed in the DSG brane setup. In order to study how this new scenario interact with the gauge field massive modes, we must obtain again a Schr\"{o}dinger-like equation. The structure of the equation (\ref{eq:susy_qm_gauge}) is held while the potential $V(z)=\gamma (\gamma\dot{A}^2+\ddot{A})$ must be revalued in terms of the DSG scenario. 

\begin{figure}
\centering
\includegraphics[width=0.99\textwidth]{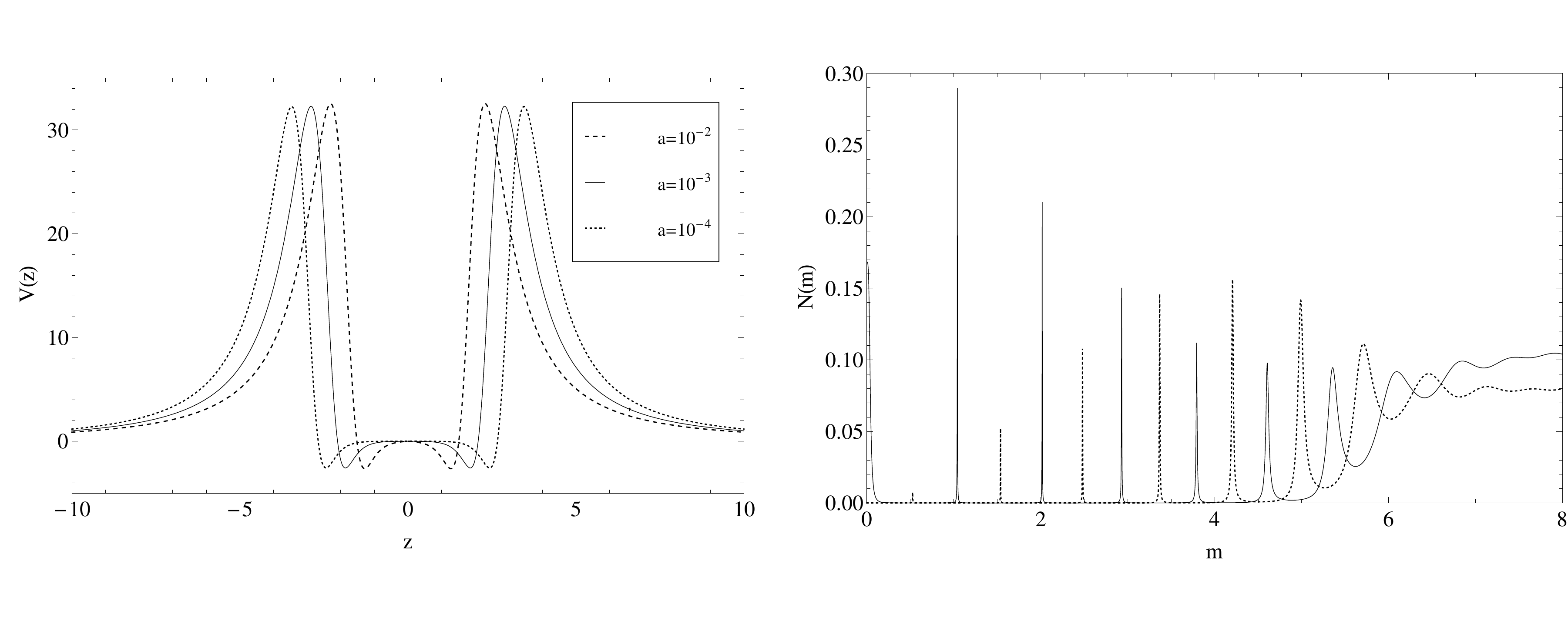}
\caption{{Plots of $V(z)$ (left) and the relative probability $N(m)$ (right) for even (solid line) and odd solutions (points). We have used $\lambda=10$ and $a=1\times10^{-5}$ in $N(m)$.}}\label{sgg7}
\end{figure}

Taking into account the results of the SG model, we know that the dilaton coupling can raise the maxima of the Schr\"{o}dinger potential. However, the influence on the potential given by the emergence of the internal structure and the splitting effect must be described. Thus, we construct $V(z)$ and show the result in the Fig. \ref{sgg7}. The potential has now two wells separated by a flat region, where $V(z)=0$.  The distance between its maximum increases with the thickness of the brane (reducing $a$). Such behaviour tells us that we must consider small values to $a$ in order to increase the probability of the wave solutions resonate.

Guided by the behaviour of Schr\"{o}dinger potential, we must examine again the massive spectrum in order to confirm the existence of resonant states. The assumption of existence of resonances is confirmed evaluating the relative probability (\ref{eq:rel_prob}). We can observe in Fig. \ref{sgg7} 
twelve resonances in $N(m)$. Such results show that the DSG 
brane is more suitable to the existence of massive modes highly coupled to the brane.

\section{Corrections to Coulomb law in the sine-Gordon branes}\label{sec-coulmb}

In order to compute the corrections from  KK gauge modes over the Coulomb law, the expression of fermion $1/2$ zero mode is required \cite{w2,hung}. Hence, we make here a brief review of fermion localization mechanism in  5D scenarios, which is already well-known in the literature \cite{nosso3, carlos, chineses6, chineses7, sg1,chineses3, chineses4, chineses5, w2}. 

Let us work in the $z$ variable and set the spin $1/2$ fermion action in 5D as \cite{nosso3, carlos, chineses6, chineses7, sg1,chineses3, chineses4, chineses5, w2}
\begin{eqnarray}\label{spin12}
S_{1/2}=\int dx^5 \sqrt{-g} \left[ \ \overline{\Psi}\Gamma^{M} D_{M}\Psi -\upsilon\overline{\Psi}\mathcal{F}(z)\Psi \ \right] ,
\end{eqnarray} 
where $\Psi=\Psi(x,z)$ is the 5D spinor, $\Gamma^{M}$ are the curved 5D gamma matrices, $D_M=\partial_M+\omega_{M}$ is the covariant derivative  with the spin connection. The scalar field $\mathcal{F}(z)$ is required to confine the zero mode \cite{nosso3, carlos, chineses6, chineses7, sg1,chineses3, chineses4, chineses5, w2} and $\upsilon$ is a coupling constant.

The Clifford algebra in our sign convention reads to $\left\{\Gamma^{M},\Gamma^{N}\right\}=+2g^{MN}$. The vielbeins are given by $g_{{MN}}=\e^{M}_{\bar{M}}\e^{M}_{\bar{N}}\eta_{{\bar{M} \bar{N}}}$. Thus, $\Gamma^M=\e^{-A(z)}\gamma^{M}$, where $\gamma^{M}$ are the 4D usual gamma matrices. The non–vanishing
terms of spin connection in our metric ansatz yield to $\omega_{\mu}=\frac{1}{2}\dot{A}(z)\gamma_{\mu}\gamma_{5}$. From this, we obtain the motion equation of action Eq. \eqref{spin12} as
\begin{eqnarray}\label{eqm-spin-12}
\left[\gamma^{\mu}\partial_{\mu}+\gamma^{5}\left(\partial_z+2 \dot{A}(z)\right)-\upsilon\e^{A(z)}\mathcal{F}(z)\right]\Psi(x,z)=0\ .
\end{eqnarray} 
At this point we define KK chiral decomposition for spin $1/2$ field as
\begin{eqnarray}\label{dec-spin-12}
\Psi(x,z)=\e^{-2A(z)}\sum_n \left[\psi_R^{(n)}(x)R_{n}(z)+\psi_L^{(n)}(x)L_{n}(z)\right].
\end{eqnarray} 
with $\gamma^{5}\psi_{R,L}(x)=\pm\psi_{R,L}(x)$. 

After applying the decomposition  \eqref{dec-spin-12} in the equation of motion \eqref{eqm-spin-12} the  chiral decoupled equation reads to the two independent Schr\"{o}dinger-like equations below
\begin{eqnarray}\label{sch-spin-12}
\left[-\partial_z^2+V_{R}(z)\right]R_n(z)=m_n^2R_n(z), \qquad \left[-\partial_z^2+V_{L}(z)\right]L_n(z)=m_n^2L_n(z) \ ,
\end{eqnarray} 
where 
\begin{eqnarray}\label{pot-spin-12}
V_{R,L}(z)=\left(\upsilon\mathcal{F}(z)\e^{A(z)}\right)^2\pm \upsilon\e^{A}\left[\mathcal{F}(z)\dot{A}(z)+ \dot{\mathcal{F}}(z)A(z)\right] \ .
\end{eqnarray}
 
The following orthogonal conditions are imposed over the functions $L(z)$ and $R(z)$
\begin{eqnarray}\label{ort-spin-12}
\int_{-\infty}^{\infty}{dz}R_nR_m=\delta_{m,n}= \int_{-\infty}^{\infty}{dz}L_nL_m, \quad \int_{-\infty}^{\infty}{dz}R_nL_m=0\ ,
\end{eqnarray} 
and the massless mode of Eq. \eqref{sch-spin-12} reads to
\begin{eqnarray}\label{0-spin-12}
L_0(z)=\zeta_{f}\e^{-\upsilon\int_0^{z}{dz'}{\mathcal{F}(z')\e^{A(z')}}}, \quad R_0(z)=\zeta_{f}\e^{+\upsilon \int_0^{z}{dz'}{\mathcal{F}(z')\e^{A(z')}}} \ ,
\end{eqnarray} 
where $\zeta_{f}$ is a normalization constant for massless fermion modes. Note that the expressions of $L_0(z)$ and $R_0(z)$ differ only by the signs in their exponentials. Hence,  it is not possible to confine both the massless right and left handed fermion modes simultaneously.

\begin{figure}[!htb] 
\begin{minipage}[t]{0.49 \linewidth}
        \centering
\includegraphics[width=0.99\textwidth]{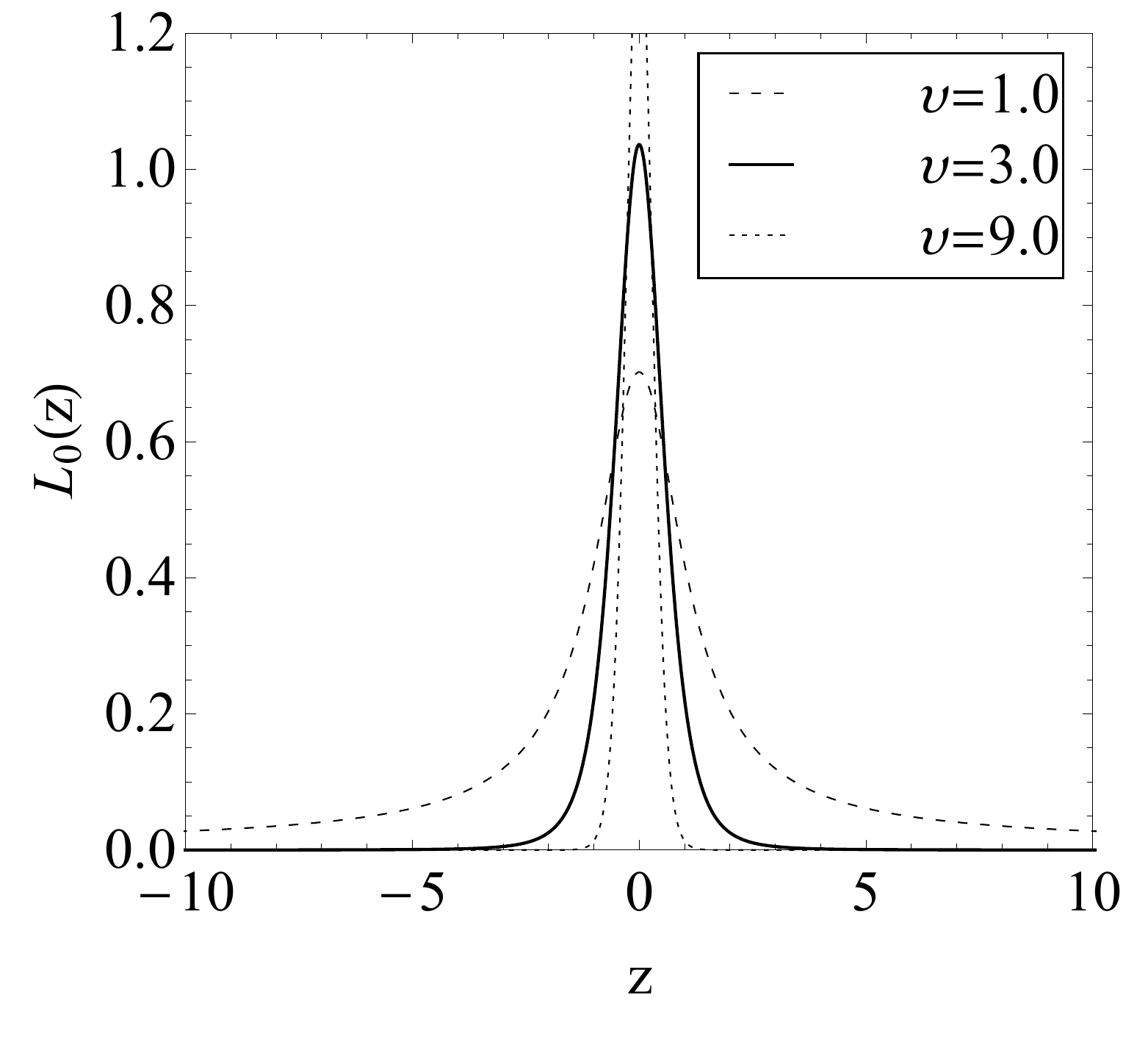}
\caption{$L_0(z)$ normalized massless left handed modes for some $\upsilon$ values in the SG scenario. These functions are localized and represent a $\delta(z)$ distribution when $\upsilon \to \infty$. The parameters $c=1$ and $b=\frac{3}{4}$ were used.}
\label{fig-L0}
\end{minipage}
\quad
\begin{minipage}[t]{0.49 \linewidth}
\includegraphics[width=0.99\textwidth]{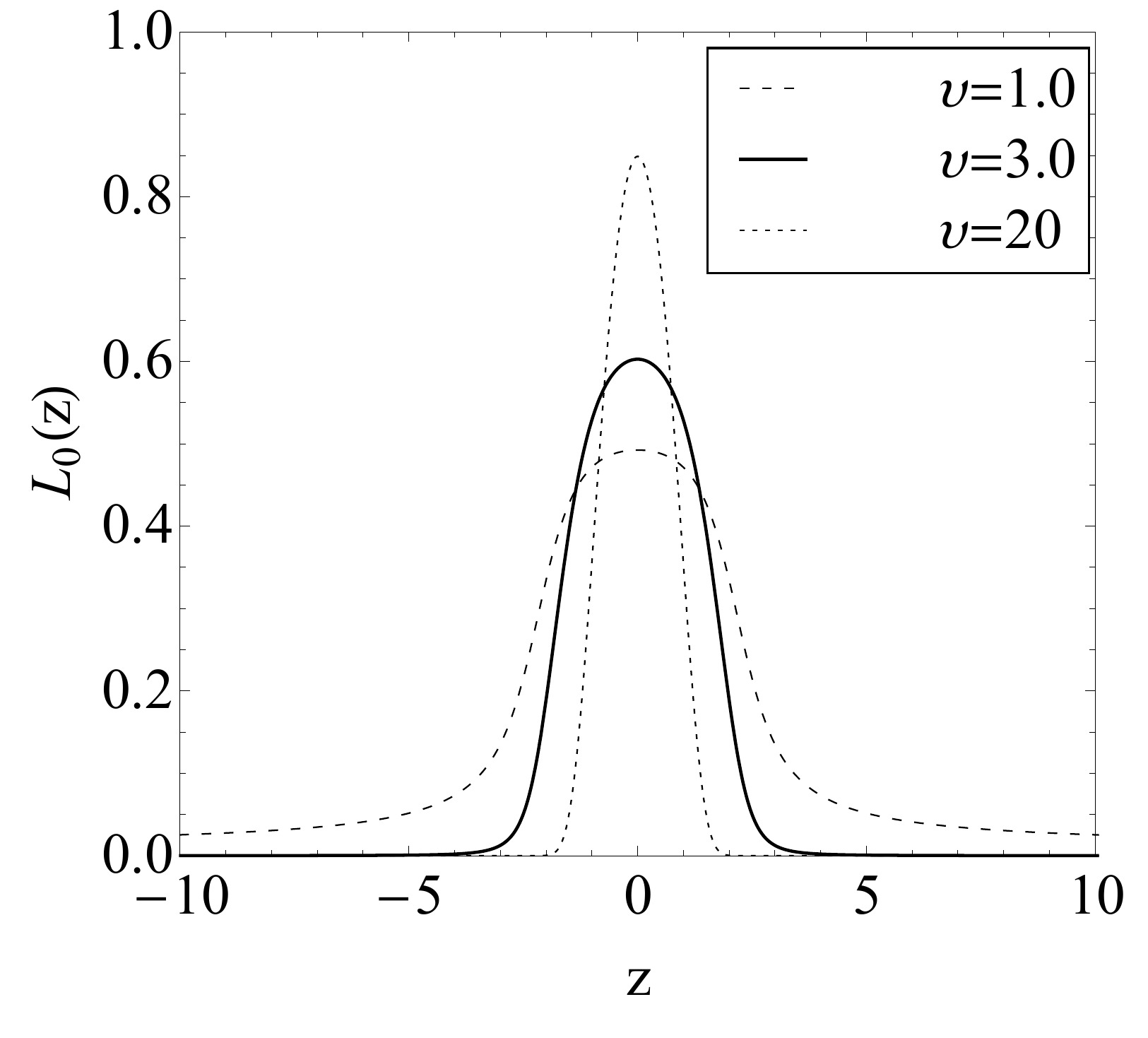}
\caption{$L_0(z)$ normalized massless left handed modes for some $\upsilon$ values in the DSG scenario. These functions are localized and represent a $\delta(z)$ distribution when $\upsilon \to \infty$. The parameters $c=1$ and $b=\frac{3}{4}$ were used.}
\label{fig-L0b}
\end{minipage}
\end{figure}

From now on, we impose for both SG and DSG models $\mathcal{F}(z)=\phi(z)$ \cite{sg1}. So, only the left handed mode will be localized in the brane. The plot of this $L_0(z)$ is presented in Fig. \ref{fig-L0} for the sine-Gordon model and in Fig. \ref{fig-L0b} for the Double sine-Gordon model.
 
Now, we focus in the physical contribution of the massive gauge modes presented in equation \eqref{schgauge} which modifies the Coulomb law at short distances in four dimensional spacetime. As a matter of fact, it was proposed in Ref. \cite{w2} a potential created by the Yukawa interaction between two fermions and a gauge field with the action in the form
\begin{eqnarray}\label{si}
S_{I}=\int dx^5 \sqrt{-g}\ \left(-e_5\right)\overline{\Psi}(x,z)\Gamma^{M} A_{M}(x,z)\Psi(x,z)
\end{eqnarray} 
where $e_5$  is a coupling constant.

From the gauge  decomposition in Eq. \eqref{sep} with the transformation in Eq. \eqref{t} and the fermion decomposition in Eq. \eqref{dec-spin-12} (where only the massless left handed mode alive) we have the above action as
\begin{eqnarray}
S_{I}=\sum_{n=0}^{\infty}\left(-e_5\right)\int{dx^4}\left(\overline{\psi}_L^{(0)}(x)\gamma^{\mu}a_\mu^{(n)}(x)\psi_L^{(0)}(x)\right)\int{dz} L^2_0(z)\e^{-\frac{(\lambda+1)}{2}A(z)} \overline{U}_n(z). \
\end{eqnarray} 
We can split the summatory into massless and massive indexes as
\begin{eqnarray}\label{c2}
S_{I}=\int{dx^4}\left[-\hat{e} \ \overline{\psi}_L^{(0)}(x)\gamma^{\mu}(x)a_\mu^{(0)}\psi_L^{(0)}(x)\ - \ \sum_{n=1}^{\infty}\epsilon_n\overline{\psi}_L^{(0)}(x)\gamma^{\mu}a_\mu^{(n)}(x)\psi_L^{(0)}(x)\right], \
\end{eqnarray} 
where the charges are
\begin{eqnarray}
\hat{e}=\e_5\int{dz} L_0^{2}(z)\e^{-\frac{(\lambda+1)}{2}A(z)}\overline{U}_0(z)=\zeta_g \e_5 \ , \label{c3} \\
\epsilon_n=\e_5\int{dz} L_0^{2}(z)\e^{-\frac{(\lambda+1)}{2}A(z)}\overline{U}_n(z)\label{c4}. \
\end{eqnarray} 

The $\hat{e}$ is the usual 4D charge of the fermion trapped on the brane. The $\zeta_g$ is the normalization constant of gauge zero mode obtained from the Eq. \eqref{schgauge} as  $\overline{U}_0(z)=\zeta_g\e^{+\frac{(\lambda+1)}{2}A(z)}$. So, by Eq. \eqref{c3} we have the charge $\e_5=\hat{e}\zeta^{-1}_g$. These expressions are similar to those presented in Ref. \cite{w2}, but modified by the dilaton parameter $\lambda$.

Finally, the correction over the Coulomb law due to massive gauge modes and due to the massless left handed fermion mode  in the limit that $n$ is continuum converges to
\begin{eqnarray}
V_c(\x)=\frac{\hat{e}^2}{4\pi \x}+\int_{m_0}^{\infty}dm\frac{\epsilon^2_n}{4\pi \x}\e^{-m \x}
=\frac{\hat{e}^2}{4\pi \x}\left[1+\Delta V_c(\x)\right] \ , \label{c5}
\end{eqnarray}
where $\x$ is the modulus of distance in 4D, $m_0$ is the first excited gauge massive mode (given by the value of squared root of quantum analogue potential  maximum in Eq. \eqref{schgauge}) and the $\Delta V_c(\x)$ is given by
\begin{eqnarray}\label{deltav}
\Delta V_c(\x)={\zeta}^{-2}_{g}\int_{m_0}^{\infty}{dm}\e^{-m \x}\left(\int_{-\infty}^{\infty}{dz}L_0^{2}(z)\e^{-\frac{(\lambda+1)}{2}A(z)} \overline{U}_n(z)\right)^2 \ .
\end{eqnarray} 

Therefore, we conclude that the 5D massless gauge mode is responsible to reproduce the 4D Coulomb potential (see Eq. \eqref{c5}), while the massive modes perform a small deviation in this potential by  the $\Delta V_c(\x)$ term. In the integration over the variable $z$ in Eq. \eqref{deltav}, the $L_0(z)$ is the only term that converges at infinity, while $\e^{-\frac{(\lambda+1)}{2}A(z)}$ grows exponentially and the $\overline{U}_n(z)$ are oscillating. Hence, in order to obtain a finite integration, the term $L_0(z)$ need to be dominant.

\begin{figure}[!htb] 
\begin{minipage}[t]{0.45 \linewidth}
        \centering
                \includegraphics[width=\linewidth]{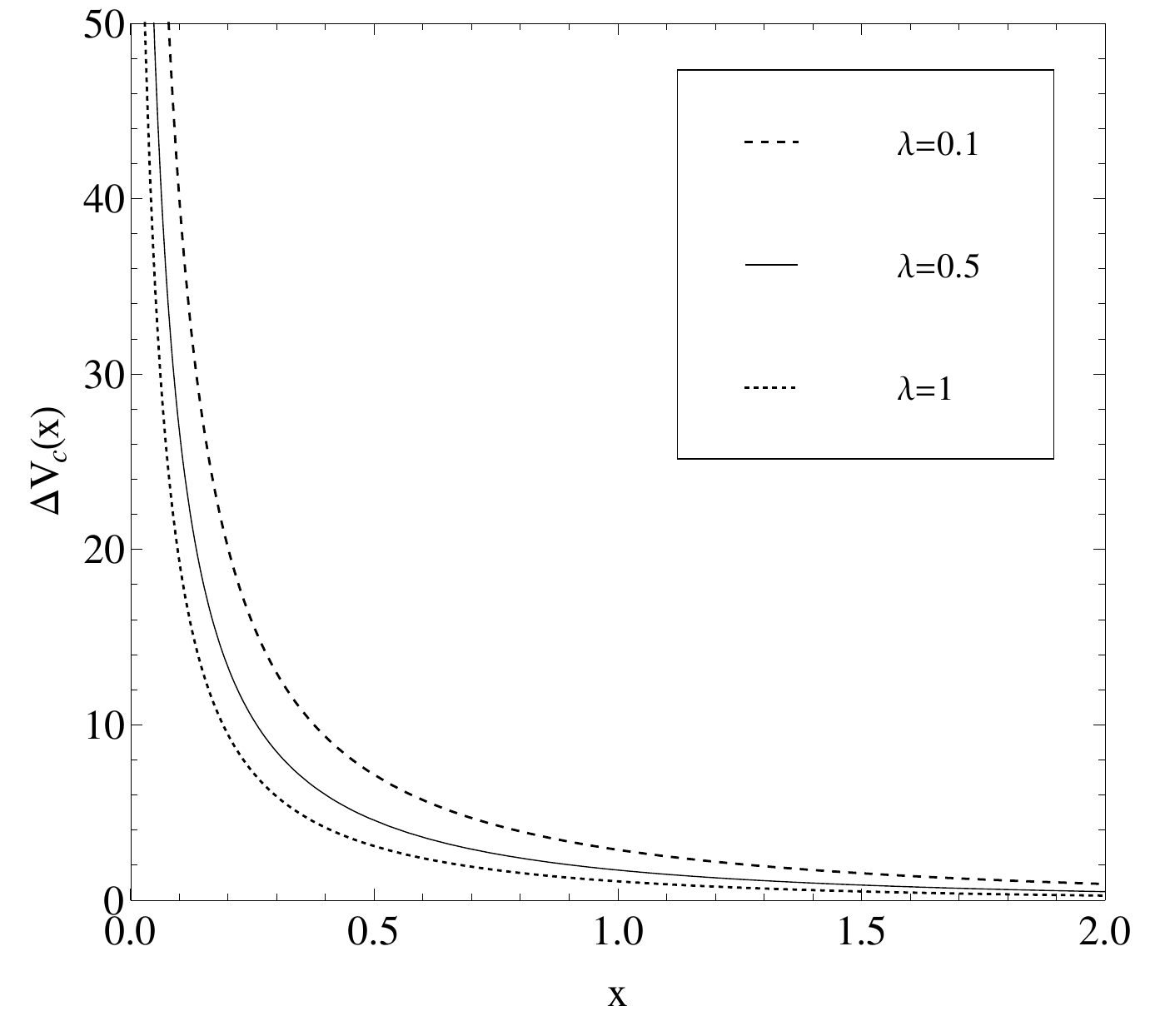}
                \caption{$\Delta V_c(\x)$ in SG model for some values of dilaton parameter $\lambda$. The parameters $\upsilon=1000$, $c=1$, $b=\frac{4}{3}$. For the three  $\lambda=\{0.1; \ 0.5; \ 1.0\}$ their respective first masses are $m_0=\{0.449; \ 0.577; \ 0.738\}$. The last mass value is set as $m=100$ for the numerical integration.}
\label{fig-deltav-sg}
\end{minipage}
\quad
\begin{minipage}[t]{0.45 \linewidth}
                \includegraphics[width=\linewidth]{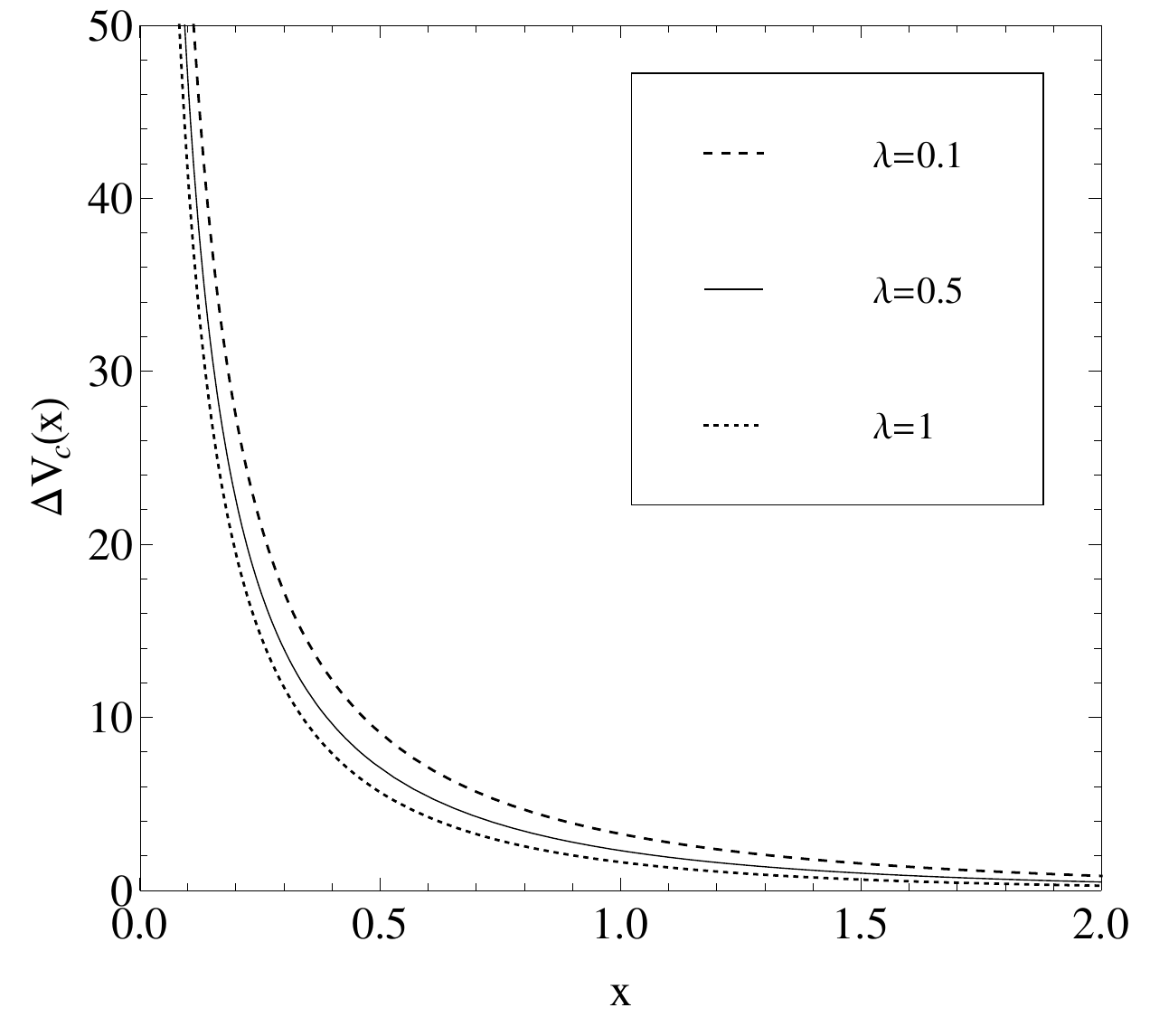}
                \caption{  $\Delta V_c(\x)$ in DSG model for some values of dilaton parameter $\lambda$.  The parameters $\upsilon=1000$, $c=1$, $b=\frac{4}{3}$. For the three  $\lambda=\{0.1; \  0.5; \ 1.0\}$ their respective first masses are $m_0=\{0.680; \ 0.876; \ 1.122\}$. The last mass value is set as $m=100$ for the numerical integration.}
                \label{fig-deltav-dsg}
\end{minipage}
\end{figure}

We perform the numerical evaluation of $\Delta V_c (x)$ and this plot is presented in Fig. \eqref{fig-deltav-sg} and in Fig. \eqref{fig-deltav-dsg} for sine-Gordon model and Double sine-Gordon model, respectively. It is important to point out that only the gauge massive even solutions contribute to these corrections, since that for the odd  solutions the integration in the Eq. \eqref{deltav} is always null for a symmetric interval. Moreover, in the case where fermions coupling constant  $\upsilon \to \infty$ the function $L^2(z)\sim\delta(z)$  filter only the  $\overline{U}_n(0)=0$ for all odd solutions.

On the other hand, from Fig. \eqref{fig-deltav-sg} and from Fig. \eqref{fig-deltav-dsg} we conclude that in both models, high  $\lambda$ values  imply the faster decay of $\Delta_c V(\x)$. The SG scenario has corrections more closed to origin than those for the same parameter in the DSG scenario. For higher value of $\upsilon$ fermions coupling constant, the correction reads to $\Delta_c V(\x)\propto \frac{\e^{-m_0\x}}{\x}$, which is also verified in Ref. \cite{w2}. Moreover, this correction is dominated by the sector
of small massive vector modes. The potential yields to  $V_c(\x)\approx\frac{\hat{e}^2}{4\pi \x}\left[1+\alpha\frac{\e^{-m_0\x}}{\x}\right]$, (where $\alpha$ is a proportionality constant) and this correction in the Coulomb potential decreases exponentially with the distances $\x$, as also verified in the corrections to Newton law in the 5D braneworlds \cite{w3}. However, the limit of Coulomb law is experimentally validated for very small distances, at least $10^{-18}$ m \cite{w4}, hence this correction should be in fact very small. 

\section{Conclusions \label{conclu}}

We have analysed the localization of the vector gauge field in three brane models generated by the sine-Gordon potentials. Our first attempt shows that the original SG braneworld is not capable of supporting gauge field zero modes. The two solutions to the gauge field in the fifth dimension lead to divergent effective actions. Analysing the massive modes we transform the gauge field equation in a Schr\"{o}dinger-like equation, however, it cannot be written in the form corresponding to supersymmetric quantum mechanics and we cannot exclude the possibility of tachyonic states in the spectrum. The potential of the Schr\"{o}dinger-like equation does not have the usual volcano-like structure tending to infinity with negative values. Such behaviour suggests that there are no resonant states to the gauge field for the SG setup. This suspect is confirmed when we evaluate the relative probability function in terms of the mass. Some peaks were found on the evaluation of the $N(m)$ function, however, the solutions of the equation of motion to the respective masses show us no reverberations to the wave solution in this scenario. 

Motivated by previous results in the literature concerning the gauge field localization, we have included the dilaton field in the SG scenario. The dilaton coupling does not changed the structure of the warp factor, but it allows the existence of a gauge field zero mode localized on the brane. Better results were also found in the analysis of the massive modes. The resulting Schr\"{o}dinger-like equation excludes the existence of tachyonic states, and the corresponding potential presented a volcano-like structure indicating the possibility of having resonant states.

The SG setup reveals us the existence of a resonant mode at $m=0.0999432$. We observe clearly that the introduction of the dilaton field to the SG setup was essential to the localization of the zero mode and the appearance of reverberations. The peak presented in the density probability, the Fig. \ref{sgg5} and the corresponding solution $U(z)$, show a massive mode with high probability to be found on the brane.

Furthermore, a new brane scenario generated by a double sine-Gordon scalar field potential was introduced.  The DSG potential revealed a new topology and resulted in a brane composed by two kinks with a splitting effect. In this new scenario, beyond the existence of a localized bound state, we have the presence of multiresonances. The Schr\"{o}dinger potential presents a structure that can be adjusted in terms of the dilaton coupling and the brane thickness, which could be appropriate to host resonant states. The results of the exploration of the massive spectrum reveal us that the DSG brane is compatible with the presence of multi quasi-localized massive states.  

Finally, we verify  how the massive gauge modes affect the Coulomb potential in 4D for both SG and DSG scenarios. These corrections are very slight and exponentially suppressed for small distances in 4D, as expected. As higher the value of the dilaton parameter, more closed to the origin are placed these corrections. The suppression of these corrections in the SG scenario is comparatively larger than in the DSG for the same set of parameter values.


\section*{Acknowledgments}
The authors thank the Coordenação de Aperfeiçoamento de Pessoal de Nível Superior (CAPES), the Conselho Nacional de Desenvolvimento Científico e Tecnológico (CNPq, grant numbers PQ 305766/2012-0, UNIVERSAL  448142/2014-7, PQ 305678/2015-9, PQ 305853/2013-9), and Fundação Cearense de apoio ao Desenvolvimento Científico e Tecnológico (FUNCAP) for financial support.


\end{document}